\documentclass[pre,twocolumn,floatfix,aps]{revtex4}
\usepackage{amsmath}
\usepackage{makeidx}
\usepackage{amssymb}
\usepackage{graphicx}

\usepackage[usenames]{color}
\usepackage[normalem]{ulem}

\usepackage{hyperref}
\usepackage[T1]{fontenc} 

\begin{document}

\title{Spatial order in a two-dimensional spin-orbit-coupled spin-1/2 condensate: superlattice, multi-ring and stripe formation }

\author{S. K. Adhikari\footnote{sk.adhikari@unesp.br      \\  https://professores.ift.unesp.br/sk.adhikari/ }}

\affiliation{Instituto de F\'{\i}sica Te\'orica, Universidade Estadual Paulista - UNESP, 01.140-070 S\~ao Paulo, S\~ao Paulo, Brazil}
      

\date{\today}

\begin{abstract}

We demonstrate  the formation of  stable spatially-ordered states   
in a  {\it  uniform} and also {\it trapped}  quasi-two-dimensional (quasi-2D) Rashba or  Dresselhaus  spin-orbit (SO) coupled pseudo 
spin-1/2 
Bose-Einstein condensate    using  the mean-field Gross-Pitaevskii equation. 
For weak SO coupling,   one  can have  a circularly-symmetric 
 $(0,+1)$- or $(0,-1)$-type  multi-ring state  with intrinsic vorticity,
for Rashba or Dresselhaus SO coupling, respectively,
where the numbers in the parentheses  denote the net  angular momentum projection in the two components,  in addition to  a circularly-asymmetric degenerate state with zero net angular momentum projection.  
For intermediate  SO couplings, in addition to the above two types,  one  can also have  states with stripe pattern in component densities with no periodic modulation in total density. 
The stripe state continues to exist for large SO coupling. In addition, a new spatially-periodic state appears in the uniform system: a {\it superlattice} state, possessing some properties of  a {\it supersolid},   with  a square-lattice pattern in component densities and also  in total density.   In a trapped system the superlattice  state is slightly different with multi-ring pattern in component density and a square-lattice pattern in total density. 
For an equal mixture of  Rashba and Dresselhaus SO couplings, in both uniform and trapped systems, only stripe states are found for all strengths of SO couplings.  In a uniform system all these states are quasi-2D solitonic states.

\end{abstract}

 \maketitle
 

\section{Introduction}

The pursuit of a supersolid  \cite{sprsld} has lately gained impetus among research workers in
different areas of  low-temperature physics. 
A supersolid is a special form of matter possessing the properties of both a crystalline solid
and a superfluid.  It is 
characterized by a spatially-ordered periodic 
matter resulting from a breakdown of continuous translational invariance as in a crystalline solid.  It also enjoys a frictionless flow as in  a superfluid, breaking continuous gauge invariance. Hence a supersolid is a special type of superfluid and its search has been confined to a superfluid, e.g.,   
Bose–Einstein condensate (BEC) \cite{bec}, and Fermi superfluid \cite{fermi}. 
The search of a supersolid helium \cite{4} was
inconclusive \cite{5}.  
Following theoretical suggestions to create a supersolid
with finite-range \cite{8} and 
dipolar \cite{7} interactions, different experimental groups confirmed supersolidity 
 in a  quasi-one-dimensional (quasi-1D) \cite{10}  and  a quasi-two-dimensional
(quasi-2D) \cite{9} dipolar trapped  BEC.

 There have been theoretical suggestions for creating supersolid-like states in a spin-orbit-coupled (SO-coupled) BEC \cite{13}.
Although there cannot be a natural spin-orbit (SO) coupling in a neutral atom, an artificial synthetic  SO coupling is possible  by tuned   
Raman lasers that couple the different spin component states 
\cite{thso}  of a    
spinor BEC \cite{exptspinor}. 
Two such possible SO couplings are due to   Rashba \cite{SOras} and Dresselhaus \cite{SOdre}. 
An equal mixture of these SO couplings has been experimentally realized 
in a pseudo spin-$1/2$ ($F=1/2$)  $^{23}$Na \cite{na-solid} and 
$^{87}$Rb  \cite{exptso}
BEC of  $F_z=0,-1$
spin component states.  
A pseudo spin-1/2 state contains only the  components  $F_z=0,-1$ of a hyperfine spin-1 ($F=1$) state. The mean-field equation of this system is governed by the Pauli spin matrices and hence the name pseudo spin-1/2.  
 Later,   
an SO-coupled 
 spin-1 $^{87}$Rb BEC of $F_z =\pm1,0$ spin component states was  also realized  \cite{exptsp1}.
Recently, a spatially-periodic  supersolid-like state with stripe pattern in density,
called a   superstripe state,  was observed  in an SO-coupled quasi-1D 
pseudo spin-1/2 spinor BEC of $^{23}$Na atoms \cite{14}  employing an equal mixture of Rashba and Dresselhaus couplings. {There have been theoretical studies on the stability of different phases of 
 BECs with Rashba-Dresselhaus spin-orbit coupling against quantum and thermal fluctuations \cite{baym}.}

In view of the observed superstripe state in an  SO-coupled  quasi-1D  pseudo spin-1/2 spinor BEC \cite{14},
in this paper we look for a supersolid-like state \cite{sprsld} in a quasi-2D  pseudo spin-1/2 SO-coupled spinor BEC using the mean-field Gross-Pitaevskii (GP) equation \cite{r6}. Previous considerations were limited to a stripe state in a  quasi-1D  pseudo spin-1/2 SO-coupled spinor BEC \cite{2020,stripe}. 
In a quasi-2D trap, different types of spatially-ordered  2D lattice states   seem likely in an  SO-coupled  pseudo spin-1/2 spinor BEC, specially after the confirmation of such states in an SO-coupled quasi-2D spin-1 spinor BEC \cite{adhisol,adhitrap}. 
We will consider both uniform  (trapless) and trapped states in this study under the action of Rashba or Dresselhaus SO coupling, in addition to an equal mixture of Rashba and Dresselhaus couplings.

First we will consider a uniform system, where the localized state  appears in the form of a quasi-2D soliton.   In a scalar BEC, solitons cannot be stabilized in two  \cite{townes} and three dimensions  \cite{r1} due to a collapse instability.     However, 
a pseudo spin-1/2 SO-coupled  BEC can sustain a quasi-1D \cite{quasi-1d} or a quasi-2D \cite{quasi-2d} or a   3D  \cite{quasi-3d} soliton.
 We identify a variety of solitons in an SO-coupled quasi-2D pseudo spin-1/2 spinor BEC.  For a weak SO coupling there can be two types of degenerate solitonic states with very large spatial extension: circularly-symmetric $(0,\pm 1)$-type multi-ring and circularly-asymmetric solitons, where the numbers in parenthesis indicate the  angular momentum projection in the two components with the upper (lower) sign corresponding to Rashba (Dresselhaus) SO coupling. The $(0,\pm 1)$-type multi-ring state has  1/2 unit of total angular momentum projection and often is classified as a half-vortex state \cite{half-vortex,hpu}. 
Each component of the circularly-asymmetric soliton hosts an antivortex-vortex pair resulting in zero net angular momentum projection in each. With the increase of SO coupling, a stripe state with stripe pattern in component densities 
appears.  For strong SO coupling two types of degenerate solitonic states appear: a state with stripe pattern in density without any modulation in total density and a  superlattice state with square-lattice pattern in both component and total densities.  The present  
solitons with a   2D square-lattice structure in total density \cite{2020}, viz. figures \ref{fig6}(d)-(f),  sharing 
properties with, and more closely related to, a conventional supersolid,  
will be termed  superlattice solitons  in the following as suggested in Refs. \cite{stripe,2020,adhisol,adhitrap}.    The multi-ring, viz. figures \ref{fig5}(a)-(c), and stripe solitons, viz. figures \ref{fig6}(a)-(c),   only exhibit a spatially-periodic pattern in the component densities
without any periodic pattern in the total density. Nevertheless, in the literature such  a state with stripe pattern in component density only, also bearing some similarity to a supersolid,   
has often been termed a superstripe state \cite{2020,stripe,14}. In this paper we call such a state by the name   stripe state to differentiate it from the superlattice state with square-lattice pattern in total density. 
For an equal mixture of Rashba     and Dresselhaus SO couplings we find only stripe solitons for all values of SO couplings; these solitons have a stripe pattern in component densities with no spatially-periodic pattern in total density.  
{
To generate an equal mixture of Rashba and Dresselhaus couplings in a laboratory a pair of Raman lasers coupling the spin-component states 
are needed \cite{exptso}, whereas to create a Rashba or a Dresselhaus coupling multiple Raman laser beams are needed
\cite{referee}, which leads to a more complex electromagnetic ``trapping'' potential.   Hence it seems reasonable that a Rashba or a Dresselhaus SO-coupled  BEC may host different types of superlattice states not possible in    the
case of an equal mixture of Rashba and Dresselhaus couplings.}

The scenario of spatially-ordered states in a harmonically-trapped SO-coupled quasi-2D 
pseudo spin-1/2 BEC remains almost the same when compared with a uniform system. 
For a weak SO coupling, the trapped system can host three types of states:  circularly-symmetric  $(0, \pm 1)$-type multi-ring state, circularly-asymmetric state and stripe state.  For large SO coupling, one encounters two types of spatially-periodic states:  stripe state and a special type of multi-ring state. This multi-ring state  possesses a multi-ring pattern in component density with a square-lattice pattern in total density near  the center  thus exhibiting a supersolid-like behavior.   

In section  \ref{i} we present the mean-field GP equation  that we use in this investigation.
In section \ref{ii}  we demonstrate the plausibility of the appearance of the $(0, \pm 1)$-type
state, stripe state, and superlattice state from a consideration of linear version of this model.
 The numerical result is presented in section III. In section \ref{a} the results for a uniform quasi-2D pseudo spin-1/2 Rashba and Dresselhaus SO-coupled  self-attractive BEC soliton are presented.  Those for an equal mixture of Rashba and Dresselhaus couplings are presented in section \ref{b}. The results for a
uniform quasi-2D pseudo spin-1/2 Rashba and Dresselhaus SO-coupled  self-repulsive BEC soliton
are presented in  section \ref{c}.  The results for  a harmonically trapped SO-coupled  quasi-2D BEC are presented in section \ref{d}.  Finally, a summary of our findings is given in section IV.

\section{Mean-field model}
\subsection{Gross-Pitaevskii equation}

\label{i}

We consider a BEC of $N$ atoms, each of mass { $m$}, under a
harmonic trap $V({\bf r})= m\omega^2(x^2+y^2)/2+ { m}\omega_z^2 z^2/2$ $(\omega_z \gg \omega)$ of  frequency $\omega_z$
and $\omega$  in the $z$ direction and  in the $x-y$ plane, respectively. The single particle Hamiltonian of the SO-coupled  BEC is
%
 \cite{exptso} 
\begin{align}\label{sp}
H_0 =-\frac { \hbar^2}{2 m}  \nabla_{\bold r}^2 +  V({\bold r})+\gamma [\eta p_y \sigma_x -  p_x \sigma_y],
\end{align}
where ${\bold r}\equiv \{x,y,z   \}$, {$\nabla_{\bold r}^2=
(\partial^2/\partial x^2+\partial^2/\partial y^2+\partial ^2 /\partial z^2) \equiv (\partial_x^2+\partial_y^2+\partial_z^2)$},
$\gamma$ is the strength of the SO-coupling term in square bracket,
{$\eta=+1$
for Rashba coupling,  $\eta =-1$ for Dresselhaus coupling, { and $\eta=0$ for an equal mixture of 
Rashba and Dresselhaus couplings,}
momentum component  $p_x=-i\hbar \partial_x, p_y=-i\hbar \partial_y$ and }
the Pauli spin matrices  $\sigma_x$ and $\sigma_y$ are
\begin{eqnarray}
\sigma_x=\begin{pmatrix}
0 & 1  \\
1 & 0 
\end{pmatrix}, \quad  \sigma_y=\begin{pmatrix}
0 & -i\\
i & 0  
\end{pmatrix}.
\end{eqnarray}

Under a strong trap in the $z$ direction, the system is assumed to be frozen in the Gaussian 
ground state in the $z$ direction and the relevant dynamics in the $x-y$ plane, { obtained by integrating out the $z$ coordinate  in a straight-forward fashion \cite{2d-3d}, leads to the   
 following set of  GP equations  at zero temperature  for spin components $F_z =  0,-1$  \cite{bao}}
\begin{align}\label{EQ1} 
i \partial_t \psi_{1}({\boldsymbol 
\rho},t)&= \left[H_{\boldsymbol \rho}
+{c_0}
n_{1}({\boldsymbol 
\rho},t) +c_2 n_2({\boldsymbol 
\rho},t) \right] \psi_{1}({\boldsymbol 
\rho},t)\nonumber \\
&+ { \gamma} (-i\eta\partial_y  + \partial_x)  \psi_{2}  ({\boldsymbol \rho},t)  \, , 
\\
\label{EQ2}
i \partial_t\psi_2({\boldsymbol \rho},t)&=\left[H_{\boldsymbol \rho}+{c_0}
n_{2} ({\boldsymbol 
\rho},t)+c_2 n_1({\boldsymbol 
\rho},t) \right] \psi_{2}({\boldsymbol 
\rho},t)\nonumber \\
&- {\gamma} (i\eta\partial_y  + \partial_x)  \psi_{1}  ({\boldsymbol \rho},t)  \, ,
  \,  
\end{align}
\begin{align}
H_{\boldsymbol \rho} &= -\textstyle \frac{1}{2}\nabla^2_{\boldsymbol \rho}+\frac{1}{2} { \rho}^2\, , \\
c_0 &= N \sqrt{2\pi\kappa}a_0, \quad c_2 = N \sqrt{2\pi\kappa}a_{12}\, ,
\end{align}
where $\boldsymbol \rho=\{x,y\}$, $\kappa = \omega_z/\omega,$  
   $\partial_t \equiv \partial/\partial t$, $\nabla^2_{\boldsymbol \rho} =\partial_x^2+\partial_y ^2$, 
 $n_j = |\psi_j|^2, j=1,2$ are the densities of spin components $F_z=  0,-1$, and $n ({\boldsymbol \rho})= \sum_j n_j({\boldsymbol \rho})$  the total density,   $a_0$  ($a_2$) is the intraspecies (interspecies) scattering length. All quantities in  (\ref{EQ1})-(\ref{EQ2}) and in the following are dimensionless; this is achieved by expressing length  in units of   oscillator length  
$l_0\equiv \sqrt{\hbar/{ m}\omega}$,
density  in units of $l_0^{-2}$, energy in units of $\hbar \omega$, and time in units of $\omega^{-1}$. 
The normalization condition is 
$ {\textstyle \int} n({x,y})\, dxdy=1\, .$

The time-independent version of 
 (\ref{EQ1})-(\ref{EQ2}), appropriate for stationary solutions, can be derived from the energy functional  
\begin{align}\label{energy}
E[\psi] &=  \textstyle  \int dx dy \big\{ \sum_j \frac{1}{2}|\nabla_{\boldsymbol \rho}\psi_j|^ 2
+\frac{1}{2}c_0(n_1^2+n_2^2)  + c_2n_{1} n_2 \nonumber \\  &
 +\textstyle\frac{1}{2}\rho ^2n + \gamma\big[  \psi_1^* (\partial_x-i\eta \partial_y )\psi_2  
- \psi_2^*(\partial_x+i\eta \partial_y)
\psi_1 \big]  \big\}.
\end{align}
   
 Equations (\ref{EQ1}), (\ref{EQ2}), and (\ref{energy}) are valid for a harmonically trapped 
quasi-2D SO-coupled pseudo spin-1/2 spinor BEC. The same for  a uniform system can be obtained by taking the limit $\omega \to 0$  and removing the harmonic trap from these equations. For a uniform system   (\ref{EQ1}) and (\ref{EQ2}) remain valid  with
\begin{align}
c_0 &= N \sqrt{2\pi}a_0, \quad c_2 = N \sqrt{2\pi}a_{12}\, .
\end{align}
 Now length is expressed in units of $l_z=\sqrt{\hbar/m\omega_z}$, density in units of $l_z^{-2}$, and energy in units of $\hbar \omega_z$, and time in units of $\omega_z^{-1}$.

\subsection{Analytic Consideration}

\label{ii}

Many properties of the density distribution can be understood from a consideration  of the single-particle Hamiltonian in the absence of nonlinear interaction. First we consider a Rashba or a Dresselhaus SO coupling. The quasi-2D  wave function of the single-particle Hamiltonian
\begin{equation}\label{abc}
H_{0}^{2D}= -\textstyle \frac{1}{2}\nabla^2_{\boldsymbol \rho}-i\gamma[\eta \partial_y \sigma_x-\partial_x \sigma_y] 
\end{equation}
 satisfies the following eigenvalue equation 
\begin{eqnarray}
\begin{bmatrix}\label{spf}
-\textstyle \frac{1}{2}\nabla^2_{\boldsymbol \rho}&  \gamma (\partial_x-i\eta \partial_y)  \\
-\gamma(\partial_x +i\eta \partial_y) & -\textstyle \frac{1}{2}\nabla^2_{\boldsymbol \rho}
\end{bmatrix}   \begin{pmatrix} \psi_1 \\ \psi_2 \end{pmatrix}
=  {\cal E}   \begin{pmatrix} \psi_1 \\ \psi_2 \end{pmatrix},
\end{eqnarray}
with energy  ${\cal E} $.
In circular coordinates $\boldsymbol \rho =\{\rho, \theta\}, x=\rho\cos \theta, y=\rho\sin\theta$, we have $(\partial_x\pm i\partial_y) = \exp (\pm i\theta) (\partial_\rho \pm i\partial_\theta /\rho),$ with $\partial_\rho \equiv \partial/\partial\rho, \partial_\theta \equiv \partial/\partial\theta$.  A circularly-symmetric solution  of angular momentum projection $m$ of
(\ref{spf}) will have the form  \cite{hpu}
\begin{eqnarray}\label{12x}
\psi_m(\boldsymbol \rho) =   \begin{pmatrix} \psi_1(\rho) \\ \psi_2(\rho) \exp (i\eta\theta) \end{pmatrix}  \frac{\exp(im\theta)}{\sqrt{2\pi}}.
\end{eqnarray}
As $\eta=\pm 1$ for Rashba and Dresselhaus SO couplings, respectively, we find that the second component carries a relative angular momentum projection $\eta$ with respect to the first component in addition to a possible overall angular momentum projection $m$.  All states with $|m|>1$ are unstable in general, hence a stable state of type (\ref{12x}) is a state of type $(0,\eta)$  or $(0,\pm 1)$ for Rashba or Dresselhaus SO coupling. Considering $m=\mp 1$, 
we can have an equivalent state of type $(\mp 1,0)$.  These two types are all states with angular momentum projection less than unity. Of these two types of equivalent states $-$ $ (0,\pm 1)$ and $(\mp 1,0)$ $-$, in this study we consider only the state 
of type   $(0,\pm 1)$. This state has angular momentum projection $m=0$ and spin $s=1/2$, hence the total angular momentum projection $j_z= m+s_z=1/2$ and this state is often called a quantum half-vortex state.  

The appearance of the stripe state can be understood from the consideration  
 that  (\ref{spf})  has the following degenerate solutions 
\begin{eqnarray}\label{eq1}
  \begin{pmatrix} \psi_1 \\ \psi_2 \end{pmatrix}=   
\begin{pmatrix} \cos(\gamma x) \\ -\sin(\gamma x) \end{pmatrix}, \\
   \begin{pmatrix} \psi_1 \\ \psi_2 \end{pmatrix}= \begin{pmatrix} \cos(\gamma y) \\ -i\eta \sin(\gamma y) \end{pmatrix}, \label{eq2}
\end{eqnarray}
with energy
\begin{align}
{\cal E} =-\frac{\gamma^2}{2}. \label{energyan}
\end{align}
The solutions (\ref{eq1}) and (\ref{eq2}) represent stripes along $y$ and $x$ directions, respectively. 
There is another set of equivalent solutions which we will not consider: $(\psi_1,\psi_2)^{\mathrm T}=  (\sin(\gamma x), \cos(\gamma x))^{\mathrm T}$  and $(\psi_1,\psi_2)^{\mathrm T}=  (\sin(\gamma y), i\eta \cos(\gamma y))^{\mathrm T}$;
these states represent stripes in density along $x$ and $y $ directions in the components. Both sets of solutions have 
uniform density in the sum of the two components: $|\psi_1|^2+|\psi_2|^2 = 1$.

\begin{figure}[!t] 
\centering
\includegraphics[width=.325\linewidth]{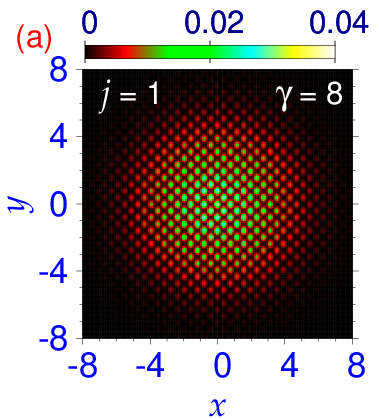} 
\includegraphics[width=.325\linewidth]{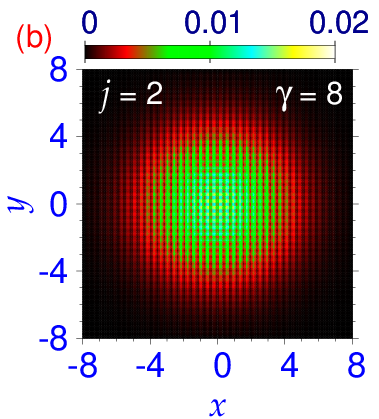}
\includegraphics[width=.325\linewidth]{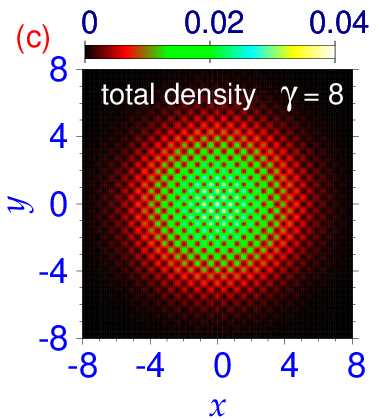}

\caption{ Contour plot of density $n_j$ of the state (\ref{eq3})  for $\gamma=8$ and $\eta =1$ 
 of components (a) $j=1$ ($n_1$), (b) $j=2$ ($n_2$) and (c) total density after multiplying  the state (\ref{eq3})   by an appropriate Gaussian distribution.  The densities are normalized as $\int dx dy n(x,y)=1$.  Results in all figures are plotted in dimensionless units.  }
\label{fig1}

\end{figure}

A linear combination of the degenerate states (\ref{eq1})  and (\ref{eq2}), e.g.
\begin{align} 
\label{eq3}
  \begin{pmatrix} \psi_1 \\ \psi_2 \end{pmatrix}=\sqrt n   
\begin{pmatrix} \cos(\gamma x) \pm  \cos(\gamma y)\\ -\sin(\gamma x)\mp i\eta \sin(\gamma y  ) \end{pmatrix},
\end{align}
is also a valid eigenfunction and represents a square-lattice pattern in density of the components as well as in the total density.  
The spatial modulation in density is produced by the $\sin$ and $\cos$ terms in  (\ref{eq3}).  

It is tempting to multiply the functions  (\ref{eq1}), (\ref{eq2}), and  (\ref{eq3}) by a localized Gaussian distribution and see if such a state can simulate the density of a localized stripe or superlattice state. The answer is affirmative. 
The density pattern of the state (\ref{eq1}) or (\ref{eq2}) after this multiplication   is quite similar to  
the density of a localized stripe state (result not shown here) obtained by a numerical solution of the GP equation.   The total density  of the stripe state does not have any spatially-periodic modulation. 
 In figure \ref{fig1}, we display a contour  plot of the density of components (a) $j=1$, (b) $j=2$, and (c) total density of the 
state (\ref{eq3}) for $\gamma =8, \eta =+1$
 after  multiplying by an appropriate Gaussian.   The density of this state is quite similar to a 
superlattice soliton with square-lattice pattern in density
for $\gamma =8, c_0=c_2=-0.5$, viz. figures \ref{fig6}(d)-(f). Even the total density in figure \ref{fig1}(c) has a square-lattice pattern as in a superlattice BEC. 
 Hence an analytic consideration of the single-particle Hamiltonian (\ref{abc}) reveals that it can naturally lead to eigenstates with densities in  agreement with those in  an  actual 
physical $(0,\pm 1)$-type state, stripe state and superlattice state in a   SO-coupled  spin-1/2 BEC. In a weakly-attractive uniform system, that we consider in this paper, the energies of these states are very close to the analytic energy ${\cal E} =-\gamma^2/2$ in most cases, indicating a negligible contribution from the nonlinear terms $c_0$ and $c_2$.

\begin{figure}[!t] 
\centering
\includegraphics[width=.95\linewidth]{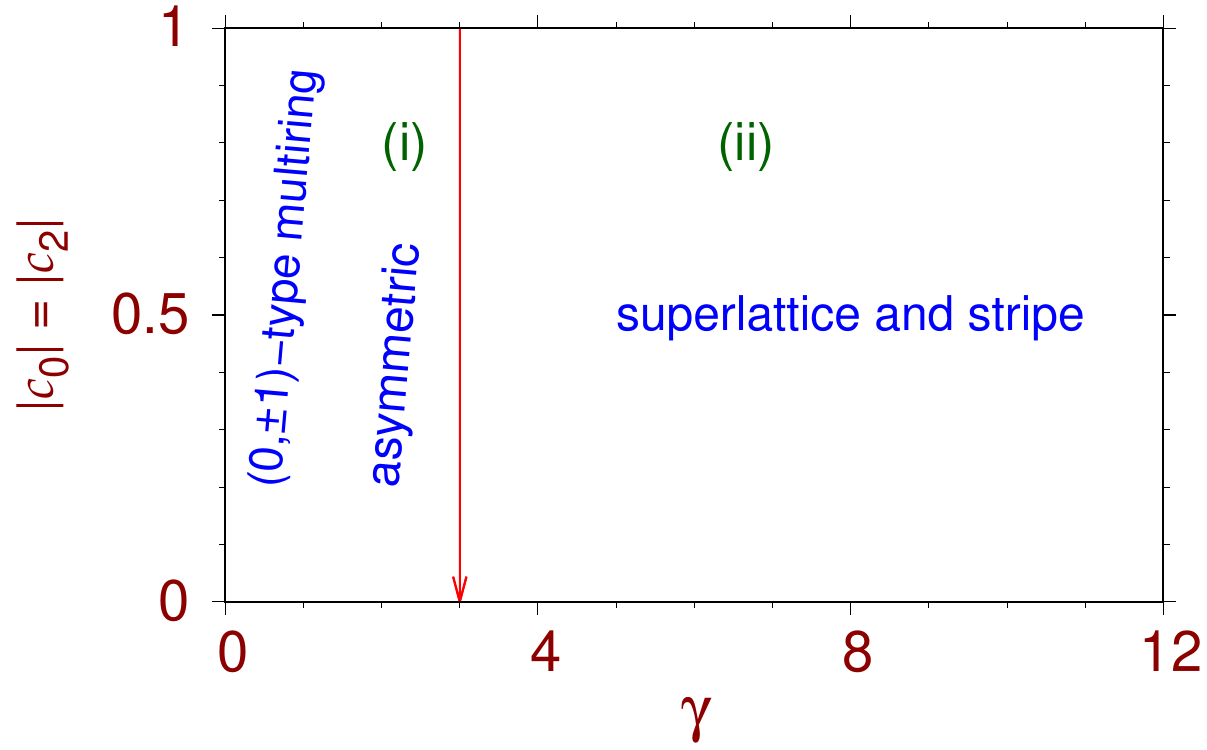}  
 
\caption{ The $c_0=c_2$ versus $\gamma$  phase plot showing { soliton formation 
in a Rashba or a Dresselhaus
SO-coupled  pseudo spin-1/2  BEC} in different regions of parameter space: (i) formation of   circularly-symmetric ($0,\pm 1$)-type
multi-ring and circularly-asymmetric solitons  and  (ii) superlattice and stripe solitons. 
 }
\label{fig2}

\end{figure}

For an equal mixture of Rashba and Dresselhaus SO couplings ($i\gamma \partial_x \sigma_y $),  (\ref{spf}) becomes 
\begin{eqnarray}
\begin{bmatrix}\label{spf2}
-\textstyle \frac{1}{2}\nabla^2_{\boldsymbol \rho}&  \gamma \partial_x  \\
-\gamma\partial_x  & -\textstyle \frac{1}{2}\nabla^2_{\boldsymbol \rho}
\end{bmatrix}   \begin{pmatrix} \psi_1 \\ \psi_2 \end{pmatrix}
= {\cal E}   \begin{pmatrix} \psi_1 \\ \psi_2 \end{pmatrix}.
\end{eqnarray}
Equation (\ref{spf2}) does not allow $(0,\pm 1)$-type state implied by  (\ref{12x}).
Equation  (\ref{spf2})  has the spatially-periodic stripe solution (\ref{eq1}).  This solution $(\cos(\gamma x), -\sin(\gamma x))^{\mathrm T}$ with energy ${\cal E} =-\gamma^2/2$ represents stripe in density along $y$ direction.  There is no degenerate solution, similar to (\ref{eq2}),  with the same energy. Thus there can be no solution of type (\ref{eq3})  in this case,  which could represent a superlattice state with square-lattice pattern in density.

\section{Numerical Result}
 
To solve  (\ref{EQ1}) and (\ref{EQ2}) numerically, we propagate
these  in time by the split-time-step Crank-Nicolson discretization scheme \cite{bec2009}
 using a space
step of $dx=dy=0.05$ and a time step of $dt=dx^2\times 0.1$ for imaginary-time propagation and  
a time step of $dt=dx^2\times 0.05$
for real-time 
propagation. { In all calculations of  stationary states, we  employ  imaginary-time approach {with the 
conservation of normalization ($=\int d {x dy [n_{1}(x,y)+n_{2}(x.y)] }$)
during time propagation,}
which 
finds the lowest-energy solution of each type. Real-time propagation is used to test the stability of the solitons.  The quantity ($M=\int d {x dy [n_{1}(x,y)-n_{2}(x,y)] }$) is
not a good quantum number and is left to freely evolve
during time propagation to attain a final converged value
consistent with the parameters of the problem. The converged value of the quantity $M$
was  found to be essentially  zero in all untrapped cases, viz. figures   \ref{fig3}-\ref{fig8},
indicating an equal number of atoms in the two components. The same could have a small nonzero value  ($\lessapprox 0.1$)   for the 
trapped case for small values of $\gamma$, viz. figure   \ref{fig9}(a)-(f).

}

\subsection{Uniform SO-coupled quasi-2D  spin-1/2 BEC: Rashba/Dresselhaus coupling}
 
\label{a}

We study the formation of  a spatially-ordered periodic pattern   in density of a self-attractive $(c_0<0)$ quasi-2D SO-coupled   uniform pseudo spin-1/2 
BEC  for different sets of parameters: the nonlinearities $c_0,  c_2$ and SO-coupling strength $\gamma$. Without losing generality we will consider only the case $c_2<0$ with $c_0=c_2$.
The scenario of soliton formation {for Rashba  or Dresselhaus SO coupling} is illustrated in the phase plot of  $c_0=c_2$ versus $\gamma$ for   $c_2 < 0$ in figure \ref{fig2} for small values of $c_0$ ($-1<c_0  <0$).  Circularly-symmetric 
$(0,\pm 1)$-type
multi-ring solitons and circularly-asymmetric solitons    are formed in region (i),  superlattice and stripe solitons are formed in region (ii).

\begin{figure}[!t] 
\centering
\includegraphics[width=.325\linewidth]{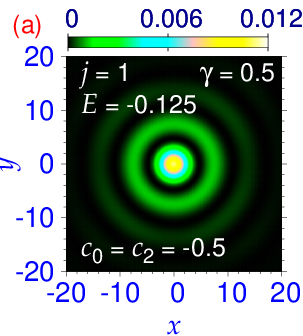} 
\includegraphics[width=.325\linewidth]{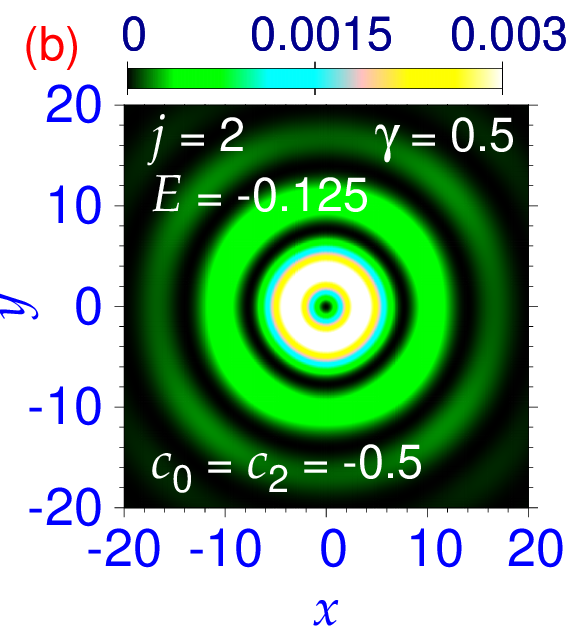}
\includegraphics[width=.325\linewidth]{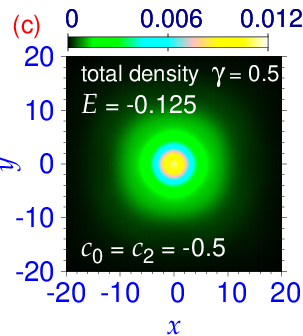}
 
 \includegraphics[width=.325\linewidth]{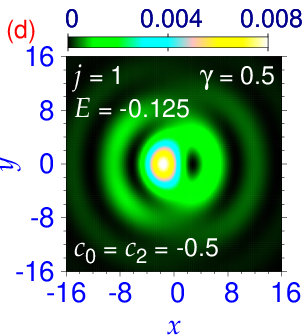} 
\includegraphics[width=.325\linewidth]{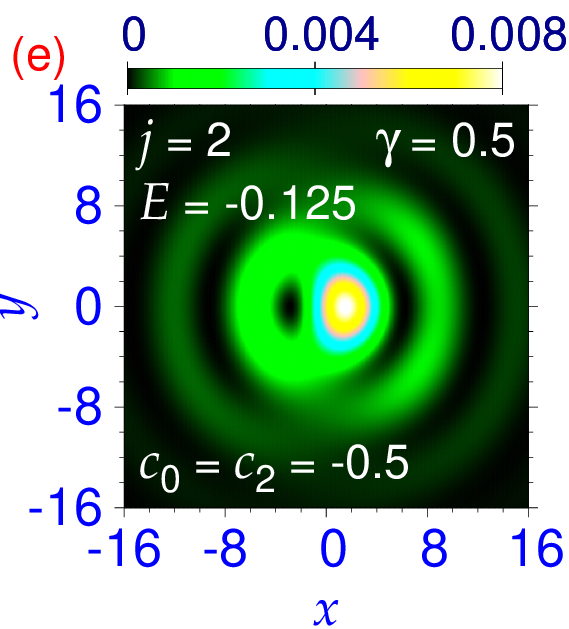}
\includegraphics[width=.325\linewidth]{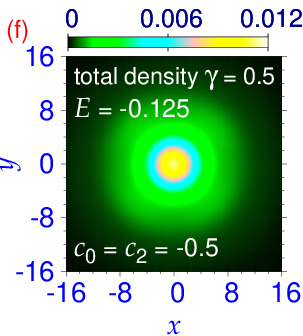}

\caption{ Contour plot of density $n_j$ of a circularly-symmetric $(0,\pm 1)$-type spin-1/2 { Rashba or Dresselhaus SO-coupled} BEC soliton 
 for components (a) $j=1$ ($n_1$), (b) $j=2$ ($n_2$) and (c) total density ($n$); the same of a circularly-asymmetric  spin-1/2 { Rashba or Dresselhaus SO-coupled} BEC soliton with an antivortex-vortex structure in each component 
 for  (d) $j=1$, (e) $j=2$ and (f) total density. The parameters are $c_0=c_2=-0.5, \gamma =0.5$.   }
\label{fig3}

\end{figure}

\begin{figure}[!t] 
\centering
\includegraphics[width=.227\linewidth]{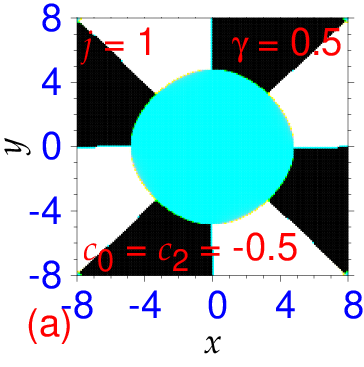} 
\includegraphics[width=.227\linewidth]{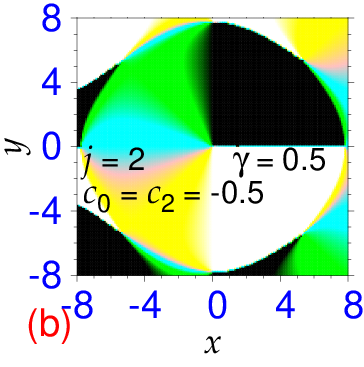}
\includegraphics[width=.227\linewidth]{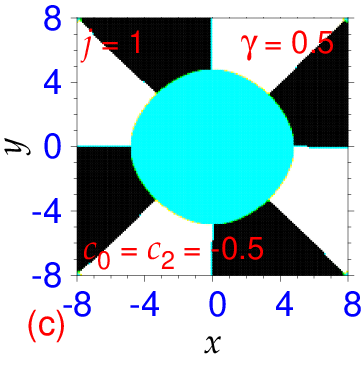}
 \includegraphics[width=.258\linewidth]{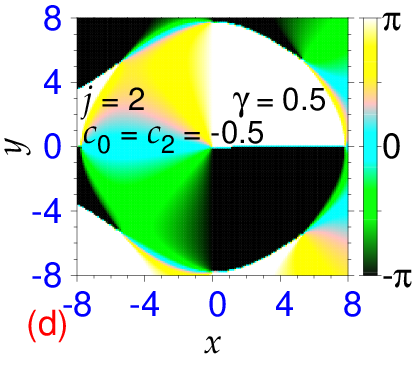} 
\includegraphics[width=.227\linewidth]{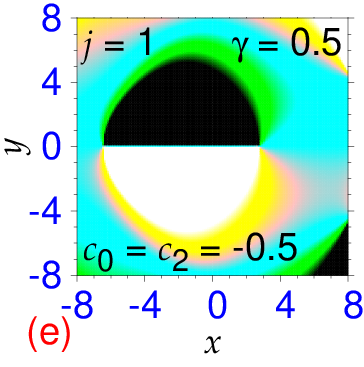} 
\includegraphics[width=.227\linewidth]{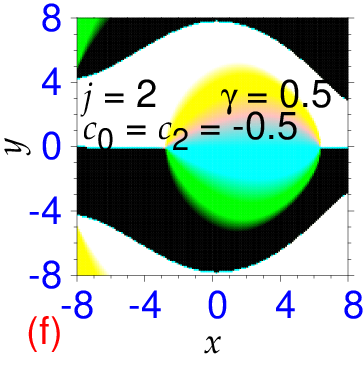} 
 \includegraphics[width=.227\linewidth]{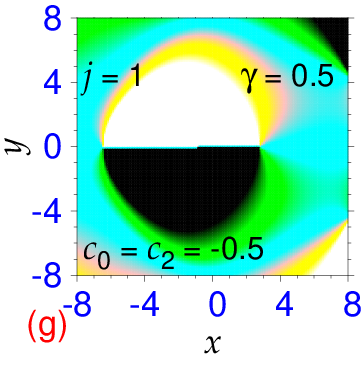} 
\includegraphics[width=.258\linewidth]{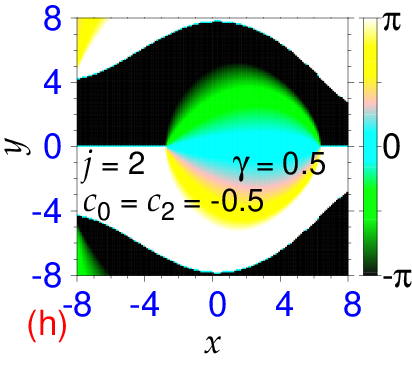}

\caption{ Contour plot of phase of  components (a) $j=1$, and (b) $j=2$ of the quasi-2D soliton of figures \ref{fig2}(a)-(b)   for  Rashba SO coupling, and  (c)-(d) the same for Dresselhaus SO coupling;
 contour plot of phase of  components (e) $j=1$, and (f) $j=2$ of the quasi-2D soliton of figures \ref{fig2}(d)-(e)   for  Rashba SO coupling, and (g)-(h) the same for Dresselhaus SO coupling. 
 }
\label{fig4}

\end{figure}

First, we consider a $(0,\pm 1)$-type multi-ring soliton for a small $\gamma$ ($\gamma = 0.5$) for Rashba  (upper sign) or Dresselhaus (lower sign) SO coupling.
In figure \ref{fig3} we display the contour plot of   density  of the central region 
of components (a) $j=1$, (b) $j=2$, and (c) total density   of a  $(0,\pm 1)$-type  multi-ring soliton. 
  In figures \ref{fig4}(a)-(b) we display the contour plot of the phase of wave function  components $j=1,2$ of the circularly-symmetric $(0,+1)$-type Rashba SO-coupled 
soliton of figures \ref{fig3}(a)-(c).  In figure  \ref{fig4}(b) there is a  phase drop of $+ 2\pi$ under a complete rotation, 
indicating an angular momentum projection of $+1$ in component $j=2$. There is no such phase drop in component $j=1$.
Although the densities of Rashba and Dresselhaus 
SO-coupled solitons are the same, the corresponding phases for Dresselhaus SO-coupling are different, viz. 
figures \ref{fig4}(c)-(d) showing the phases of components $j= 1,2$ of the Dresselhaus SO-coupled soliton. In this case in figure \ref{fig4}(d) we find a phase drop of  $- 2 \pi$  under a complete rotation,
indicating an angular momentum projection    of $- 1$  in this component. The  contour plot of density of components (a) $j=1$, (b) $j=2$, and (c) total density of a circularly-asymmetric soliton is shown in figures \ref{fig3}(d)-(f) for Rashba or Dresselhaus SO coupling. Again the phases of the two SO couplings are different. In figures \ref{fig4}(e)-(f) we plot the phase of components $j=1,2$ of the wave function of the soliton of 
figures \ref{fig3}(d)-(e) for Rashba coupling, the same for Dresselhaus coupling are shown in  figures \ref{fig4}(g)-(h).   In all cases we find an antivortex-vortex pair of angular momentum projection   $-1$ and $+1$ at two different places such that the net angular momentum projection in each component is zero. Of the vortex-antivortex pair, one is coreless \cite{coreless}. 
In imaginary-time propagation, to obtain a  $(0,\pm 1)$-type multi-ring soliton, the appropriate vortex (antivortex) for Rashba (Dresselhaus) SO coupling in component $j=2$ was imprinted in an initial  Gaussian function  as 
$\psi^{\mathrm{initial}}_2(x,y) = (x\pm iy)\times\phi_{\mathrm{Gauss}}(x,y)$.   To obtain the circularly-asymmetric soliton, the initial wave function was taken to be a localized Gaussian function in the two components.
 The numerical energy of both types of degenerate  solitons  in  figure \ref{fig3}  is $E=-0.125$, in agreement  with the analytic result ${\cal E} =-\gamma^2/2=-0.125$, viz.  (\ref{energyan}), independent of the type of SO coupling: Rashba or Dresselhaus.

\begin{figure}[!t] 
\centering

\includegraphics[width=.325\linewidth]{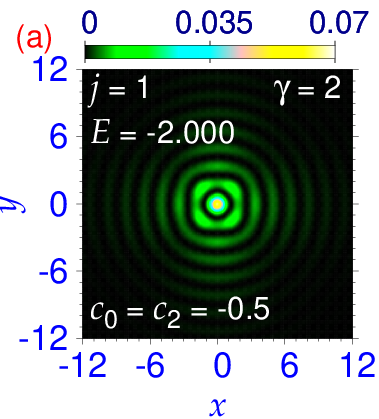}
\includegraphics[width=.325\linewidth]{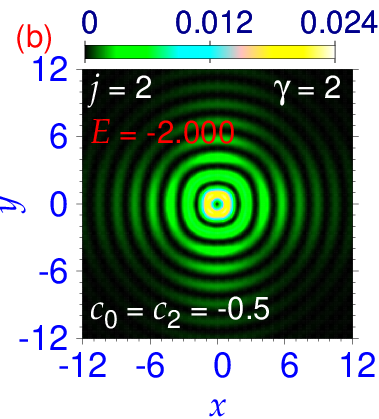}
 \includegraphics[width=.325\linewidth]{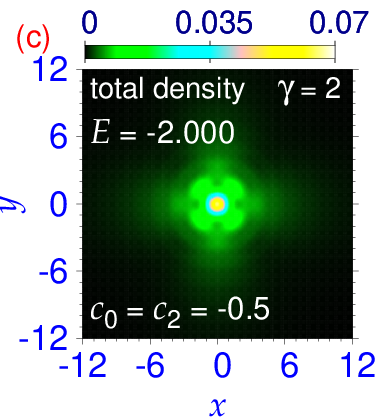} 

\includegraphics[width=.325\linewidth]{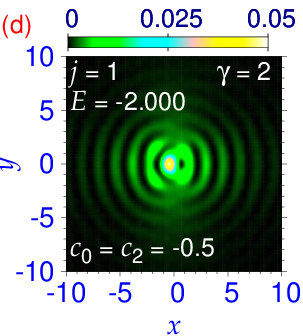}
\includegraphics[width=.325\linewidth]{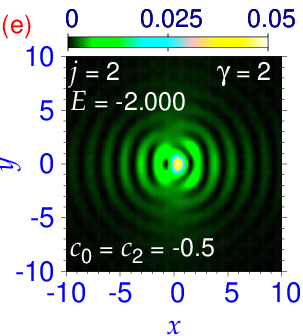} 
 \includegraphics[width=.325\linewidth]{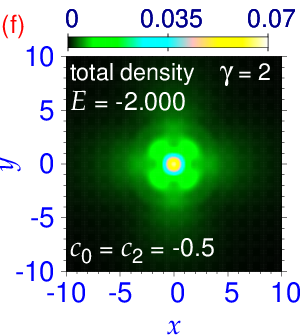} 
\includegraphics[width=.325\linewidth]{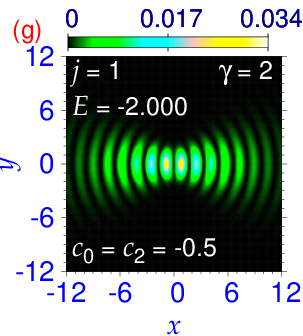}
\includegraphics[width=.325\linewidth]{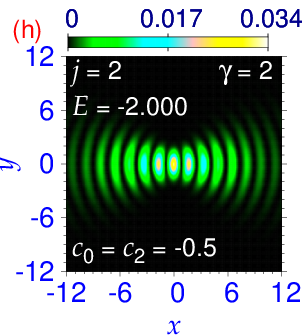}
\includegraphics[width=.325\linewidth]{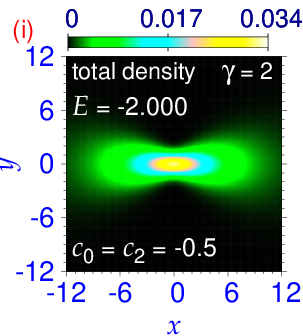} 

\caption{  Contour plot of density $n_j$ of a circularly-symmetric $(0,\pm 1)$-type spin-1/2 multi-ring { Rashba or Dresselhaus SO-coupled} BEC soliton 
 for components (a) $j=1$ ($n_1$), (b) $j=2$ ($n_2$) and (c) total density ($n$); the same of a circularly-asymmetric  spin-1/2 { Rashba or Dresselhaus SO-coupled} BEC soliton with an antivortex-vortex structure in each component 
 for  (d) $j=1$, (e) $j=2$  and (f) total density; the same of a 
spin-1/2 { Rashba or Dresselhaus SO-coupled} 
stripe soliton for  (g) $j=1$, (h) $j=2$  and (i) total density. The parameters used are 
$c_0=c_2=-0.5, \gamma=2$.}
\label{fig5}

\end{figure}

\begin{figure}[!t] 
\centering

\includegraphics[width=.325\linewidth]{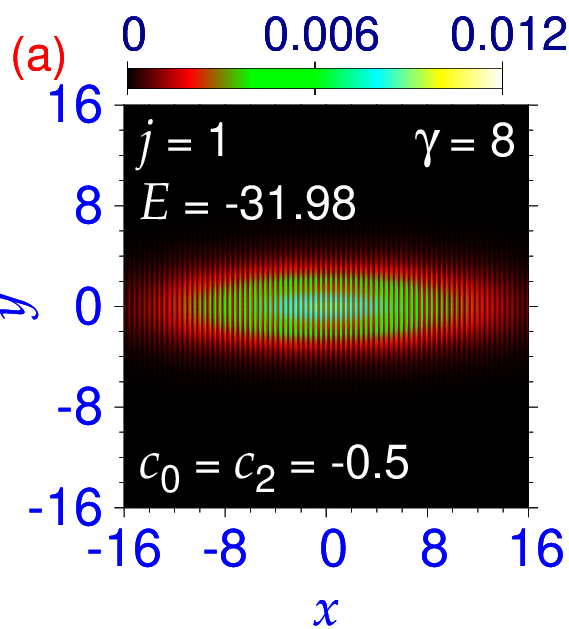}
\includegraphics[width=.325\linewidth]{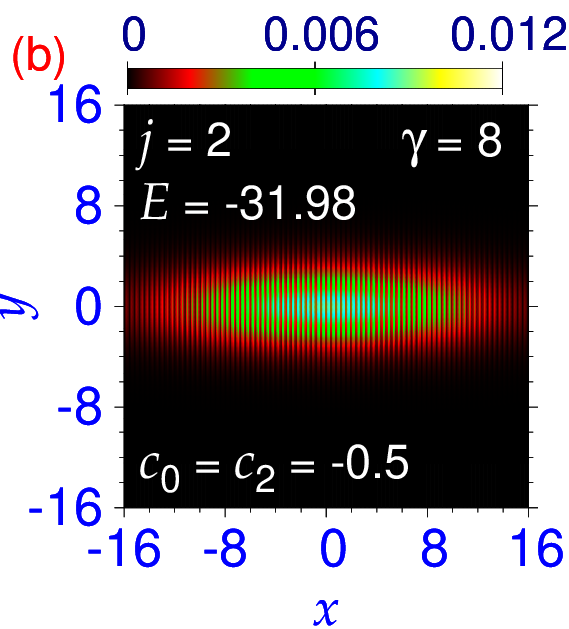}
 \includegraphics[width=.325\linewidth]{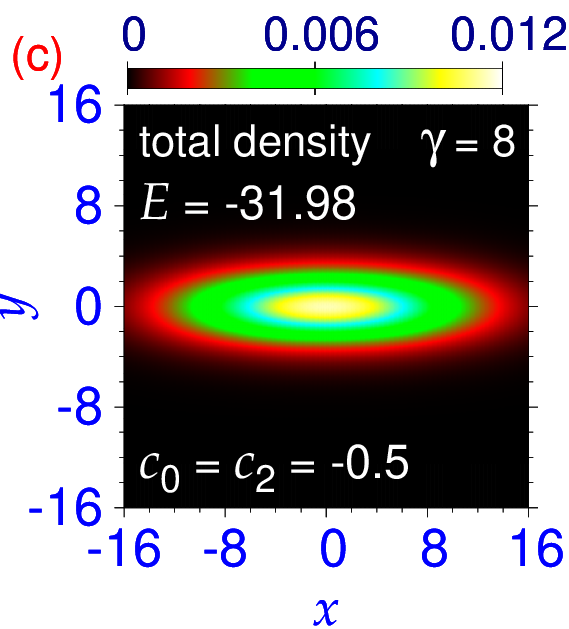} 

\includegraphics[width=.47\linewidth]{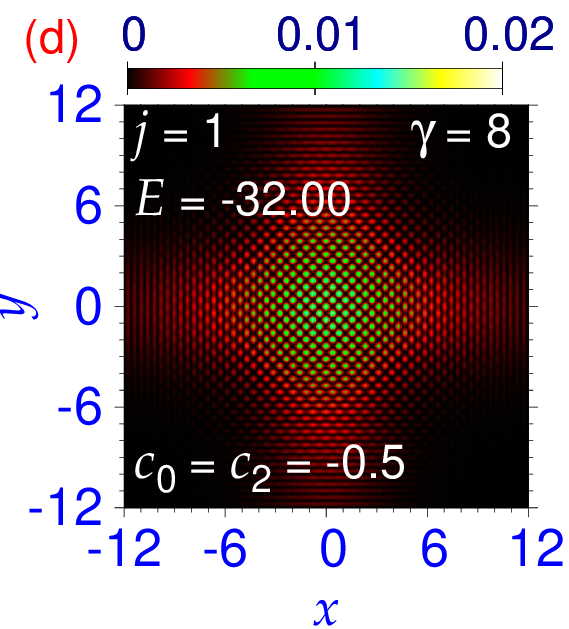}
\includegraphics[width=.47\linewidth]{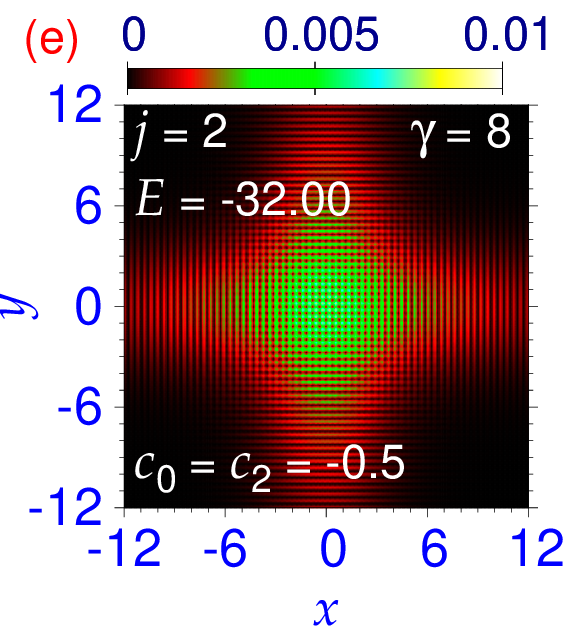} 

 \includegraphics[width=.47\linewidth]{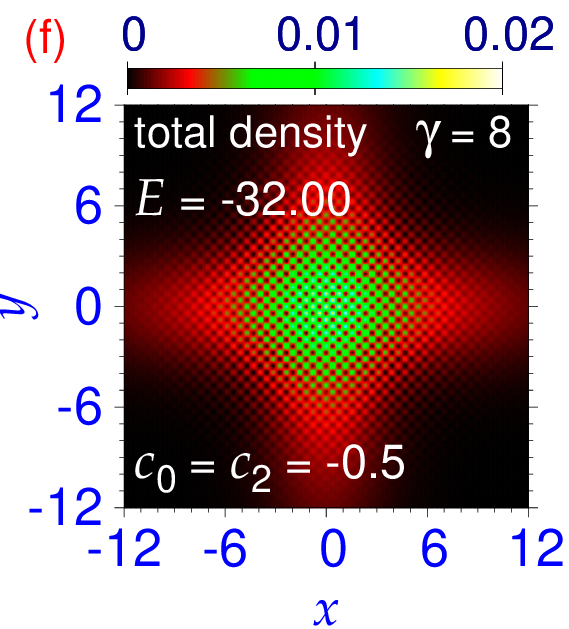}

\caption{   Contour plot of density $n_j$ of a  pseudo spin-1/2 { Rashba  or Dresselhaus SO-coupled} stripe soliton 
 of components (a) $j= 1$ and (b) $j=2$  (c) and total density.  
The same of a  superlattice  soliton 
  of components {(d) $j=  1$,  (e) $j=2$, and (f) the total density}. The parameters used are 
$c_0=c_2=-0.5, \gamma=8$.}
\label{fig6}

\end{figure}

To study   multi-ring solitons for medium  values of SO coupling ($\gamma=2$), 
in figure \ref{fig5} we show the contour plot of   density  of components (a) $j= 1$ , (b) $j=2$  
and (c) total density 
of a circularly-symmetric  $(0,\pm 1)$-type multi-ring soliton  for Rashba or Dresselhaus SO coupling
for $c_0=c_2=-0.5$.    Although,  there is a multi-ring structure  in density of  the two  components, the total density  shows no spatially-periodic modulation.  Multi-ring solitons were also investigated in a quasi-2D pseudo spin-1/2 SO-coupled BEC trapped in a radially periodic potential \cite{radper} which creates a multi-ring modulation in density. However,  the present radial modulation in density without any external trap 
is a consequence of the SO coupling. 
  The increase of $\gamma$ from figures  \ref{fig3}(a)-(c)  to  figures  \ref{fig5}(a)-(c) has increased the binding, 
and hence aids in  forming  the compact solitons. In figures  \ref{fig5}(d)-(f) we display the circularly-asymmetric  solitons  for $c_0=c_2=-0.5, \gamma =2$. These solitons are more compact and have developed asymmetric rings with the increase of $\gamma $  from figures \ref{fig3}(d)-(f) to   figures \ref{fig5}(d)-(f). In addition to these two types of solitons we find a new type of soliton not possible for a small SO  coupling ($\gamma =0.5$). These are the stripe solitons displayed in figures \ref{fig5}(g)-(i). To obtain these solitons the stripe pattern is imprinted on the initial wave function as:   $\psi^{\mathrm{initial}}_1(x,y) \sim \sin(\gamma x)\times\phi_{\mathrm{Gauss}}(x,y);  \psi^{\mathrm{initial}}_2(x,y) \sim \cos(\gamma x)$  $\times\phi_{\mathrm{Gauss}}(x,y)$. In this case the total density has no spatially-periodic modulation. The numerical energies of these three types of states for both Rashba and Dresselhaus SO couplings  are identical ($E=-2.000$) and equal to the analytic energy of  (\ref{energyan}) (${\cal E} =-\gamma^2/2=-2$); hence these states can be considered degenerate.

{As $\gamma$ is increased,  the    $(0,\pm 1)$-type
multi-ring soliton
ceases to exist and gives rise to a new type of soliton:  superlattice soliton with square-lattice modulation in density. The stripe soliton and the circularly-asymmetric soliton continue to exist.  However, as we are interested in spatially-periodic states, we will not consider the circularly-asymmetric soliton here. 
The spatial period of the lattice or stripe increases as $\gamma$ is reduced. For a small $\gamma$, the size of the soliton is smaller than this period and the periodic pattern in density  is not possible, viz figure \ref{fig3} for $\gamma=0.5$.}
 In figure \ref{fig6} we show  a quasi-2D  stripe soliton for  $c_0=c_2 =-0.5,$ and $\gamma=8$ through a contour plot of density of components (a) $j=1$, (b) $j=2$, and (c) total density, obtained by imaginary-time propagation using an initial localized wave function  modulated by appropriate stripes   in $j= 1$  and $j=2$  components. Although, there is a stripe pattern in density in this case,
the positions of maxima in component $j=1$ coincide with the minima in component $j=2$, thus resulting in a total density without modulation.

 In figure \ref{fig6}
the superlattice soliton for the same set of parameters ($c_0=c_2=-0.5, \gamma=8$)
is displayed through a contour plot of density of components (d) $j=1$, (e) $j=2$, and (f)  total density,  obtained by imaginary-time propagation using the converged wave function of figures \ref{fig5}(a)-(c) as the initial state.   
The distribution of matter on a 2D square lattice is prominent in this case, not only in the component densities but also 
in the total density, viz.
figure  \ref{fig6}(f)  \cite{2020}.
The present superlattice soliton  is a consequence of the SO coupling and breaks {\it continuous}  translational symmetry as required in a supersolid \cite{sprsld}.  
The numerical density pattern  of the components 
and the total density in this case are quite similar to the  analytic  component  density and total  density 
of figure \ref{fig1}. 
Again,
for the same  set of parameters, $c_0=c_2=-0.5, \gamma=8$,  the numerical energies of the stripe and superlattice solitons are the same   ($E=-32.00$) and in agreement with  the analytic estimate   (\ref{energyan}) (${\cal E} =-\gamma^2/2=-32$). Hence   the spatially-periodic stripe and  superlattice  solitons can be considered to be degenerate.

\subsection{Uniform SO-coupled quasi-2D  spin-1/2 BEC: Equal-mixture coupling}
 
\label{b}

\begin{figure}[!t]
\centering 
\includegraphics[width=.325\linewidth]{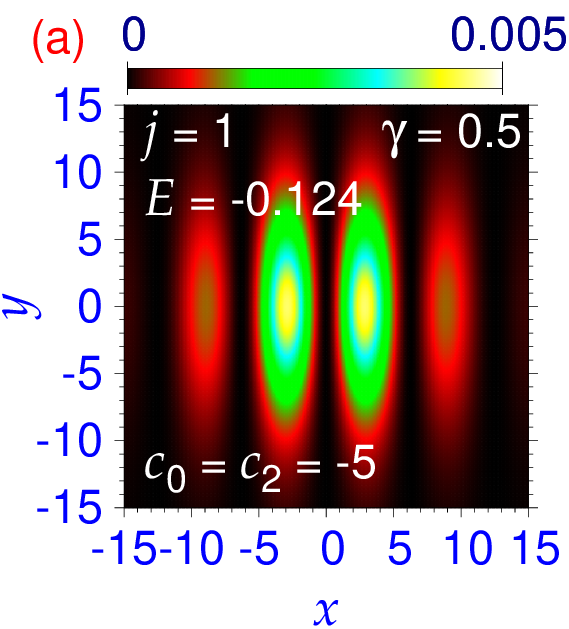}
 \includegraphics[width=.325\linewidth]{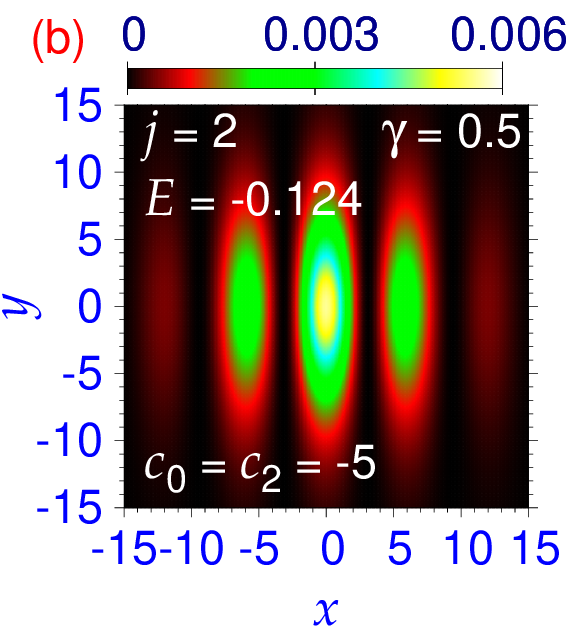} 
\includegraphics[width=.325\linewidth]{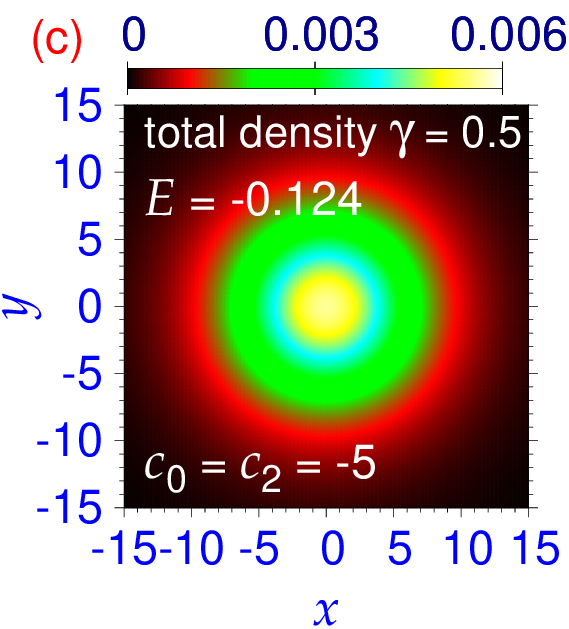}
 \includegraphics[width=.325\linewidth]{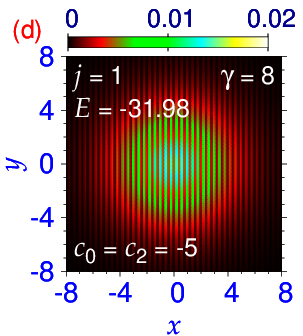}  
\includegraphics[width=.325\linewidth]{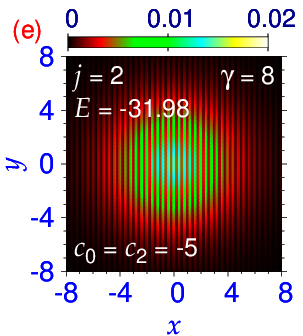}
 \includegraphics[width=.325\linewidth]{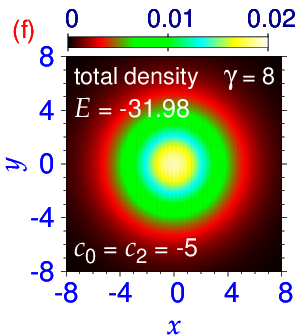}

\caption{ Contour plot of density of  a pseudo spin-1/2 stripe soliton for an equal mixture of Rashba and Dresselhaus couplings
 of components {(a) $j= 1 $, (b) $j=2$ and (c) the total density} for parameters $c_0=c_2=-5, \gamma=0.5$.
(d)-(f) The same densities for parameters $c_0=c_2=-5, \gamma=8.$}
\label{fig7}
\end{figure}

Although the Rashba or Dresselhaus SO couplings are fundamental in nature, all experiments so far employed an equal mixture of Rashba and Dresselhaus SO couplings \cite{na-solid,exptso,exptsp1}  ($-\gamma p_x \sigma_y$).  This SO coupling is simpler in nature, involving partial derivative in one direction only, and is easily realizable in an experiment.
Because of this phenomenological interest, we consider here the possibility of the formation of spatially-periodic pattern in a uniform quasi-2D pseudo spin-1/2 BEC under the action of an equal mixture of Rashba and Dresselhaus couplings. In this case, for all strengths of SO coupling $\gamma$,  one can only have a soliton with a stripe pattern in component densities  with a total density without such modulation. For a small $\gamma$  (=0.5),   the   stripe soliton     for $c_0=c_2=-5$ is 
displayed in figures \ref{fig7}(a)-(c),  where we plot the density of components $j=1$, $j=2$, and the total density. 
In this case as the derivative SO coupling is of lower dimension acting only along the $x$ direction, it has a weak localization capacity, hence we had to employ stronger attractive nonlinearity $c_0=c_2=-5$, in place of $c_0=c_2=-0.5$ used for Rashba or Dresselhaus coupling, to obtain a compact soliton.
 For large $\gamma$ (=8), 
the stripe state for $c_0=c_2=-5$, and for an equal mixture of Rashba and Dresselhaus couplings, has a circular boundary as exhibited in figures    \ref{fig7}(d)-(f) through a contour plot of densities of components $j=1$, $j=2$ and the total density.
 This boundary is elliptic  for Rashba or Dresselhaus SO coupling in   figures    \ref{fig6}(a)-(c).

\subsection{Uniform SO-coupled quasi-2D spin-1/2 self-repulsive BEC}
 
\label{c}

\begin{figure}[!t]
\centering 
\includegraphics[width=.325\linewidth]{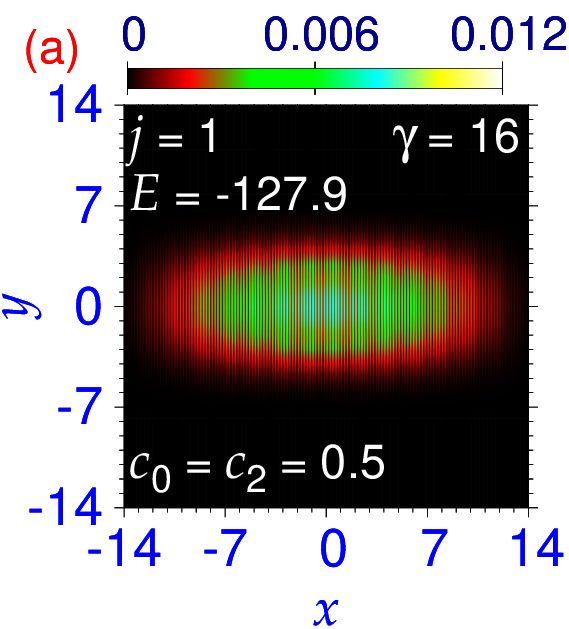}
 \includegraphics[width=.325\linewidth]{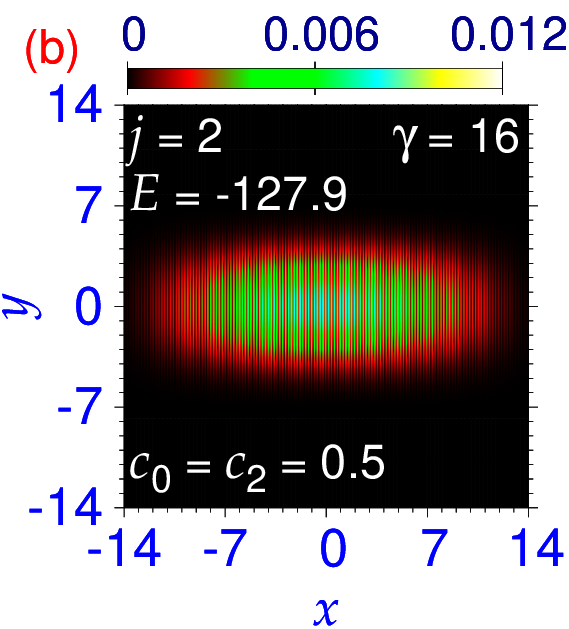} 
\includegraphics[width=.325\linewidth]{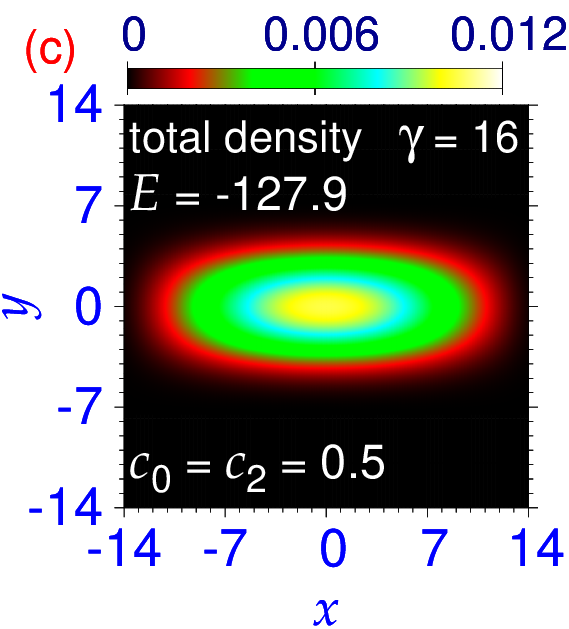}
 \includegraphics[width=.325\linewidth]{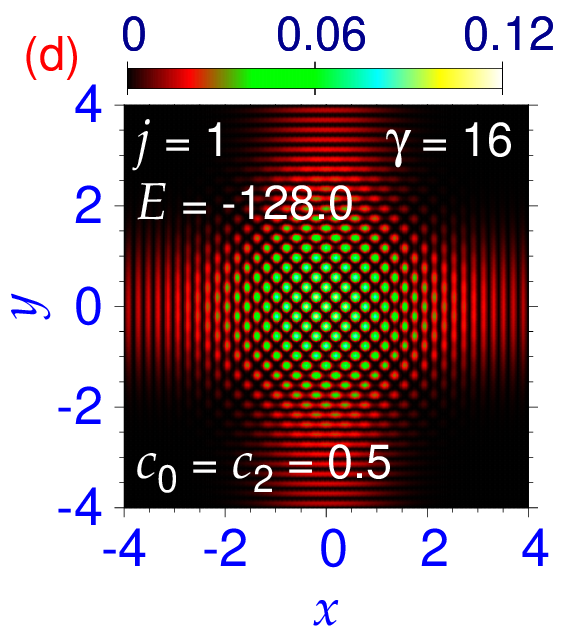}  
\includegraphics[width=.325\linewidth]{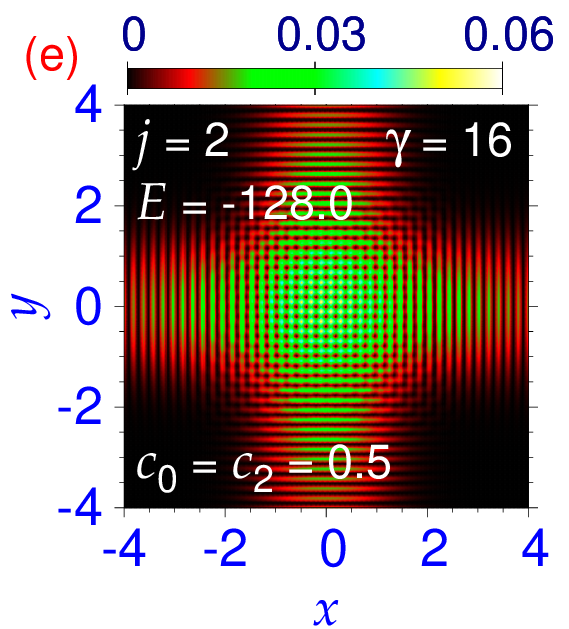}
 \includegraphics[width=.325\linewidth]{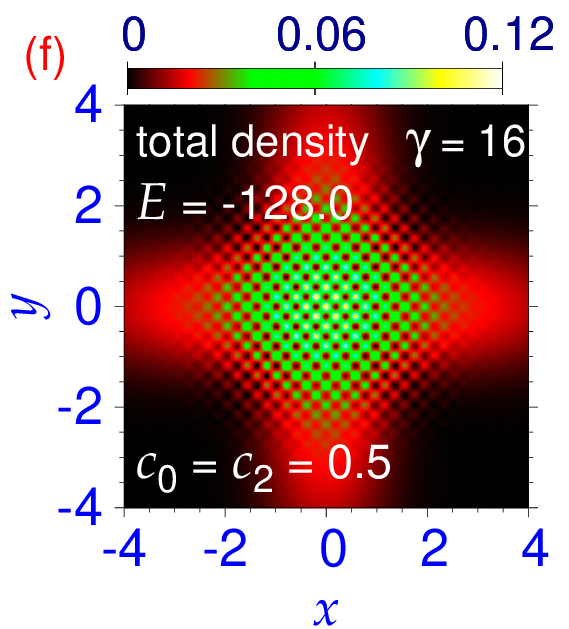}

\caption{ Contour plot of density of a self-repulsive Rashba or Dresselhaus  SO-coupled pseudo spin-1/2 stripe soliton 
 of components {(a) $j=1 $, (b) $j=2$ and (c) the total density}. 
The same of a  superlattice  soliton of components
 (d) $j= 1$,  (e) $j=2$ and (f) the total density.  The parameters employed are $c_0= c_2= 0.5, \gamma=16$.}
\label{fig8}
\end{figure}

 So far we considered a uniform SO-coupled quasi-2D self-attractive BEC ($c_0<0$).  It is also possible to have a self-repulsive ($c_0>0$) BEC. Such a self-repulsive BEC can be formed in a quasi-1D system \cite{symb}, provided that the interspecies interaction is attractive. 
In the case of a uniform SO-coupled pseudo spin-1/2 BEC, the scenario of soliton formation 
of a self-repulsive BEC with attractive interspecies interaction is the same as a self-attractive BEC. Nevertheless, due to SO coupling, one can also have a quasi-2D self-repulsive BEC soliton for repulsive interspecies interaction; such a soliton is not possible in the absence of SO coupling ($\gamma=0$).
Here we demonstrate the formation of a self-repulsive soliton in a
Rashba or Dresselhaus  SO-coupled quasi-2D pseudo spin-1/2 BEC for repulsive interspecies interaction. To this end we consider $c_0=c_2=0.5$; these values of nonlinearity  lead to compact solitons of appropriate size for a not too small $\gamma$.  In this case it is possible to have the same types of solitons as in the case of a self-attractive BEC for different strengths of SO coupling and we exhibit only the solitons for $\gamma = 16$ with pronounced spatially-periodic pattern in density. In figure \ref{fig8},  we display a contour plot of densities of an  SO-coupled quasi-2D self-repulsive stripe BEC of components  (a) $j=1$, (b) $j=2$ and (c) total density. The same of a superlattice BEC is shown in figures \ref{fig8}(d)-(f), where  the square-lattice pattern in the total density is clearly visible. The numerical energies of the stripe and the superlattice solitons  are $E=-127.9$ and $-128.0$, respectively, in agreement with the analytic estimate (\ref{energyan}) (${\cal E} = -\gamma^2/2=-128$).

\subsection{Trapped SO-coupled quasi-2D BEC: Rashba/Dresselhaus coupling}

\label{d}

\begin{figure}[!t]
\centering 
\includegraphics[width=.325\linewidth]{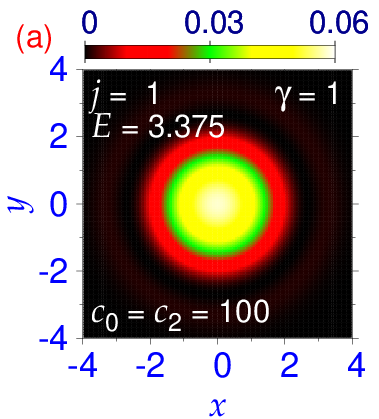}
 \includegraphics[width=.325\linewidth]{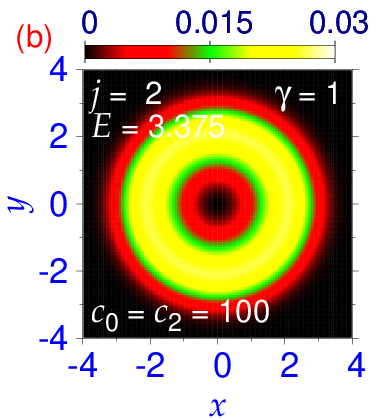} 
\includegraphics[width=.325\linewidth]{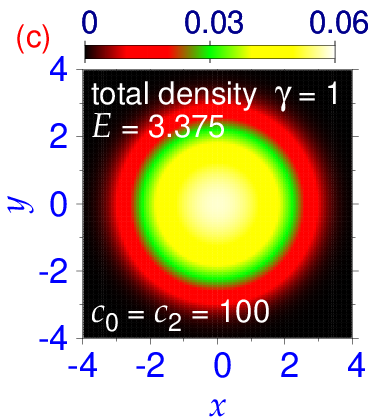}
 \includegraphics[width=.325\linewidth]{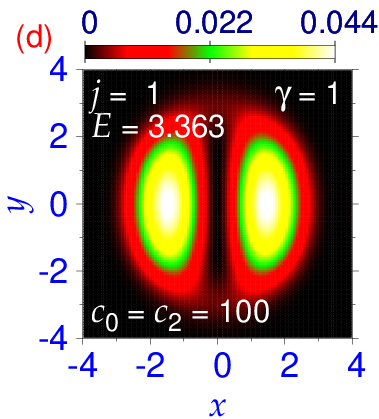}  
\includegraphics[width=.325\linewidth]{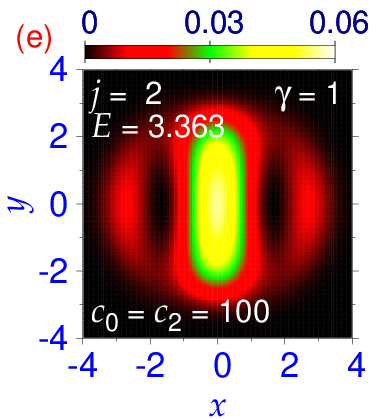}
 \includegraphics[width=.325\linewidth]{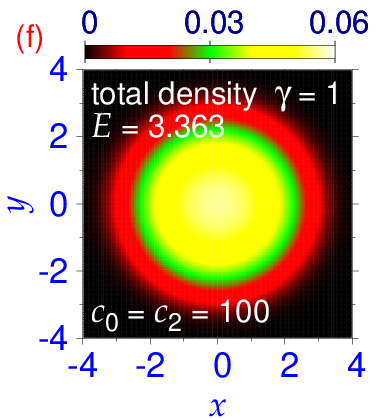}  
 \includegraphics[width=.325\linewidth]{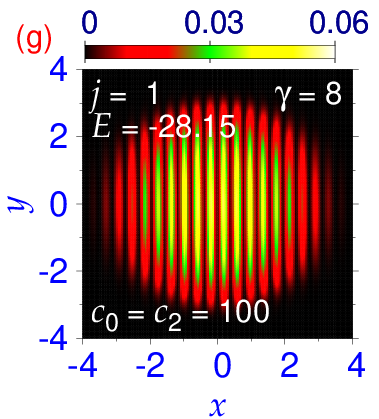}  
\includegraphics[width=.325\linewidth]{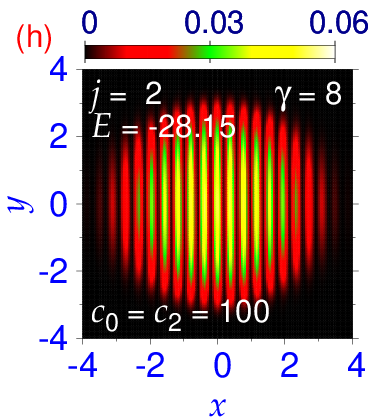}
 \includegraphics[width=.325\linewidth]{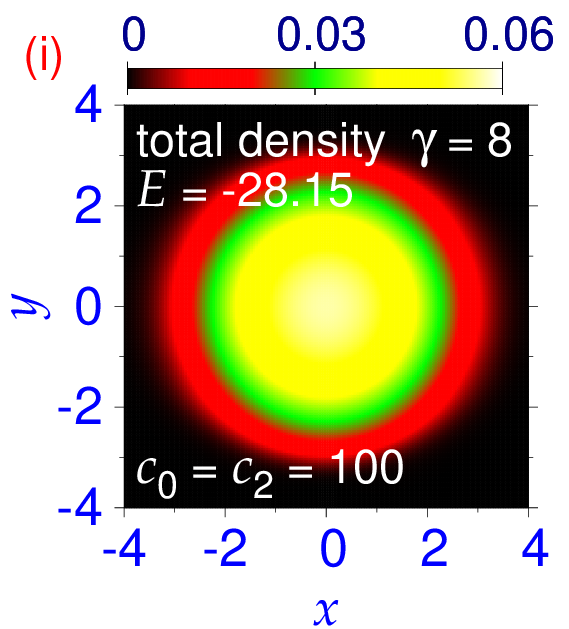}  
 \includegraphics[width=.325\linewidth]{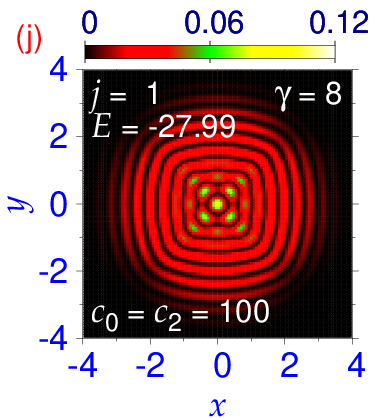}  
\includegraphics[width=.325\linewidth]{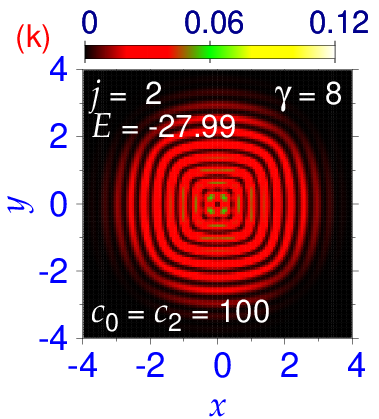}
 \includegraphics[width=.325\linewidth]{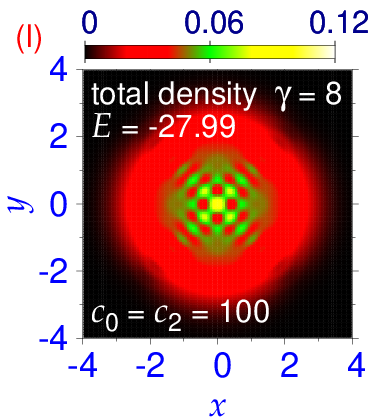}

\caption{ Contour plot of density of  a Rashba or Dresselhaus SO-coupled pseudo spin-1/2 harmonically-trapped  circularly-symmetric $(0,\pm 1)$-type  BEC 
 of components {(a) $j= 1 $, (b) $j=2$ and (c) the total density}  for $\gamma =1$.
(d)-(f) The same densities of   a stripe state.   (g)-(i)  The same densities of a stripe state for $\gamma =8$.  (j)-(l)  The same densities of a multi-ring supersolid-like  state for $\gamma =8$.  The nonlinearities are $c_0=c_2=100.$ }
\label{fig9}
\end{figure}

We now investigate the formation of a spatially-ordered state in a trapped SO-coupled quasi-2D pseudo spin-1/2 BEC for Rashba or Dresselhaus coupling.
The densities  for Rashba or Dresselhaus SO couplings are the same for all sets of parameters. 
 The scenario of the states for different strengths of SO coupling remain qualitatively the same as in the case of   a uniform system. However, there are differences for large SO-coupling $\gamma$. In this case we consider a moderate value of nonlinearity $c_0=c_2=100$. For a small $\gamma$, one can have a circularly-symmetric $(0,\pm 1)$-type state; however, due to the harmonic trap it is not possible to have a multi-ring structure in the outer region as in the case of a uniform BEC for a small $\gamma \sim 1$.  In figures \ref{fig9}(a)-(c) we display the contour plot of densities of the circularly-symmetric  $(0,\pm 1)$-type state for $\gamma=1$. 
The same for the circularly-asymmetric   state is quite similar to the densities in figures \ref{fig3}(d)-(f) (result not shown here).
The densities of the stripe state for the same set of parameters are shown in figures \ref{fig9}(d)-(f). In this case,   the circularly-symmetric   $(0,\pm 1)$-type state and the circularly-asymmetric  state 
have the same energy $E=3.375$, whereas the stripe state has a smaller energy $E=3.363$. Hence the stripe state is the ground state. In fact, it is the ground state for all strengths of SO coupling. In the uniform case, considered so far, all  these states, for  a fixed $\gamma$, had approximately the same numerical energy and hence could be considered 
degenerate.
The contour plot of densities of the stripe state for large strength of SO coupling $\gamma =8$ and for $c_0=c_2=100$ is shown in figures \ref{fig9}(g)-(i).  For small values of nonlinearity $c_0$ and $c_2$ ($c_0,c_2 \lessapprox 60$), the stripe state is the only possible stable state for large $\gamma$. For medium to large values of nonlinearity   $c_0=c_2=100$, one has a new type of excited  multi-ring state with concentric ring pattern in the density of components $j=1$ and $j=2$ as displayed in figures \ref{fig9}(j)-(l); however, the total density has a square-lattice pattern in the central region and is  without modulation in density in  the outer region.  However, the multi-ring  state  ($E=27.99$) is an excited state   compared to the stripe state ($E=28.15$), which is the ground state.  Nevertheless, the multi-ring  state has supersolid-like properties. We also studied spatially-periodic states in a trapped SO-coupled quasi-2D spin-1/2 BEC for an equal mixture of Rashba and Dresselhaus couplings (result not presented in this paper). In that case only a stripe pattern can be obtained for all strengths of SO-coupling $\gamma$.

\section{Summary and Discussion}

To search for a supersolid-like spatially-ordered state in a uniform or harmonically-trapped pseudo spin-1/2 quasi-2D SO-coupled BEC, we find spatially periodic multi-ring,  stripe,  and superlattice states for large strengths of SO coupling $\gamma$.  We consider Rashba and Dresselhaus SO couplings and an equal mixture of these couplings in this paper. For  a small $\gamma$  and for Rashba and Dresselhaus SO couplings, 
we could have two degenerate states: a circularly symmetric $(0,\pm 1)$-type multi-ring state, and a circularly asymmetric state.   
For a large $\gamma$,
the stripe state is obtained for   all types of SO couplings in both uniform and trapped BECs. The stripe state has stripe pattern in component densities and no modulation in total density. The superlattice state is found only in a uniform BEC  for Rashba or Dresselhaus SO couplings and has  a square-lattice pattern in component densities and also in total density, viz. figures \ref{fig6}(d)-(f) and \ref{fig8}(d)-(f). The multi-ring state is found in a uniform system for medium $\gamma (=2)$, viz.  figures  \ref{fig5}(a)-(c), and in a trapped system for large $\gamma (=8)$, viz. figures  \ref{fig9}(m)-(o).  In the case of a  multi-ring state in a uniform system, there is no modulation in total density, whereas in the case of a trapped multi-ring state, there is a square-lattice modulation in the total density near the center indicating supersolid-like behavior. Nevertheless, the square-lattice pattern in the superlattice states in a uniform system is more prominent than the same in the trapped multi-ring state. In a uniform system, the stripe and the superlattice states have the same numerical energy and hence they are degenerate, whereas in  a trapped system the stripe state is the ground state with lowest energy for all $\gamma$.  For an equal mixture of Rashba and Dresselhaus SO couplings, only a stripe state is obtained in all cases; 
  for a small $\gamma$,  no supersolid-like state is  obtained. For a small $\gamma$, the solitonic states in a uniform SO-coupled BEC have very large spatial extension.
  All these states are dynamically stable, as we verified by real-time simulation over a long period of time as in the case of an SO-coupled spin-1 quasi-2D BEC \cite{adhisol,adhitrap}  (result not presented in this paper).  Superlattice states were also found in an SO-coupled spin-1 spinor BEC \cite{adhisol}. However, it is more difficult to perform an experiment in an SO-coupled  spin-1 spinor BEC with three spin components compared to an SO-coupled pseudo spin-1/2 spinor BEC with two spin components. Hence we believe that the present study could motivate experiments in  the search of superlattice states in an SO-coupled  pseudo spin-1/2   spinor BEC.
  The superlattice soliton is dynamically robust and deserve   further  theoretical and experimental  investigation.

\begin{acknowledgments}
The author  
 acknowledges support by the CNPq (Brazil) grant 301324/2019-0, and by the ICTP-SAIFR-FAPESP (Brazil) grant 2016/01343-7

\end{acknowledgments}

\end{document}